\documentclass[9pt]{article}
\usepackage{epsfig}
\usepackage{graphicx}
\usepackage{amsfonts}
\newcommand{\ct}{\cite}
\begin{document}
                                                                                
\begin{center}
\Large\bf{Fibers on a graph with local load sharing}\\
\bigskip
\large\sl{Uma Divakaran \footnote{E-mail:udiva@iitk.ac.in} 
{\rm and} Amit Dutta\footnote{E-mail:dutta@iitk.ac.in}}\\
\rm{Department of Physics,\\
 Indian Institute of Technology, \\
Kanpur 208 016, India}
\end{center}
\vspace*{0.2cm}
\begin{abstract}
We study a random fiber bundle model with tips of the fibers placed 
on a graph having co-ordination number 3. These fibers follow 
local load sharing with uniformly distributed threshold strengths of the fibers.
We have studied the critical behaviour of the model numerically using a finite 
size scaling method and the mean field critical behaviour is established.
The avalanche size distribution is also found to 
exhibit a mean field nature in the asymptotic limit.
\end{abstract}
\section{Introduction}
Collapse of buildings, failure of networks and many other breakdown events
have resulted into a great deal of research to understand the cause of 
such disasters and also forced the scientists to invent measures to 
prevent such mishappenings\ct{benguigui}.
Fiber Bundle Model (FBM) is a step towards capturing the essential physics of
breakdown phenomena \ct{peirce,daniel,coleman}. 

A FBM consists of N parallel elastic fibers
with identical elastic constant but randomly distributed threshold
strengths chosen from a distribution. This random distribution of 
threshold strengths ($\sigma_{th}$) 
introduces randomness in the model and hence the model is called
Random Fiber Bundle Model (RFBM).
When the stress generated due to the externally applied force
exceeds the threshold strength of a given fiber, the fiber breaks.
The additional load due to the failure of fibers is distributed to the 
remaining intact fibers by a load sharing rule. In a Global Load Sharing (GLS)
scheme, the broken fibers redistribute their 
stress to all the intact fibers.
This is a mean field model with an effective 
long range interaction amongst the intact fibers. The critical
exponents have been obtained for such a model\ct{dynamic}.
We investigate here the other extreme load sharing rule, namely 
Local Load Sharing (LLS)\ct{pacheco, zhang, wu}. In a local load 
sharing rule, the broken fiber gives its load only to the 
nearest intact fibers. Therefore, this load sharing depends upon the 
dimensionality of the system concerned. As the external force is increased, more
fibers break.  
The redistribution of the stress of a broken fiber for a given applied load 
takes place until a fixed point is reached when no more failures take place. 
If the external load exceeds a particular 
load called $critical~ load$, 
the complete breakdown of the bundle takes 
place. The determination of this critical load and the study of the associated
phase transition are of prime interest.

It is well established
that the fibers with LLS in a one-dimensional 
regular lattice do not show any critical behaviour and the critical
stress $\sigma_c$ (=critical load per fiber) 
goes to zero in the thermodynamic limit\ct{pacheco}. Models with
load sharing rule interpolating between GLS and LLS have also been studied
\ct{herrmann}.
Kim, Kim and Jeong\ct{complex} studied fibers placed on different network 
models: namely, the  Erdos-Renyi model
of random network, Watts-Strogatz network and the static model of a scale free
network with LLS. They established the mean field (GLS) critical behaviour of
 fiber bundles residing on
these networks. In Ref.\ct{moreno01} the instability introduced in a large 
scale free network by the triggering of node-breaking avalanches is
analyzed using fiber bundle model as the framework.
It should be noted that FBM can be related to a fuse model where breakdown
of a fuse is caused due to a large applied voltage and the system is eventually
driven to a completely broken state as the external voltage is increased 
beyond the critical voltage. The breaking profiles of such a model
 has also been studied on a scale 
free network\ct{pinheiro}. Under certain load modes, this breaking profile 
resembles the stress-strain curve of the dynamics of the fibers.

In this paper we look at the dynamics of fibers following local
load sharing rule when placed on a graph with co-ordination number 3. 
We will point out how our model is different from the above 
models later in this paper.

\section{The model}

Dhar, Shukla and 
Sethna \ct{dhar} studied a
Random Field Ising Model on a Bethe lattice
using a particular random graph\ct{barabasi} for simulating the Bethe lattice
structure.  
We have used the same approach to generate a geometry which has 
connections as described below.
\begin{figure}
\begin{center}
\includegraphics[height=1.8in,width=2.1in]{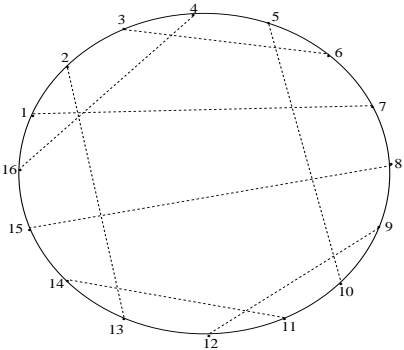}
\end{center}
\caption{An example of a graph with coordination number 3 and 16 sites.}
\end{figure}

\noindent $\bullet$ Label the sites by integers from 
1 to $N$ where $N$ is even. 

\noindent $\bullet$ Connect site $i$ to $i+1$ for all $i$ 
such that site $N$ is connected to site 1 (periodic boundary condition). 
We call this connection a direct connection. Thus we have a ring of $N$ sites.

\noindent $\bullet$ Now connect each site $i$ randomly to a unique site 
other than $i+1$ and $i-1$ thus forming $N/2$ pairs of sites. 
This connection is called 
random connection. 

Random connection is unique for a given site. For example, in Fig.~1 where
we show a particular configuration, site 
number $14$ is directly connected to $13$, $15$ and randomly connected to $11$
which is unique for site number $14$. At the same time, random connection of
site $11$ then becomes $14$. 
Thus any site will have two directly connected
neighbours and one unique random neighbour making the co-ordination
number of each site to be three at the beginning of the dynamics.
In this construction, all sites are on the same footing. The graph resembles
a Newman-Watts (NW) model where nearest neighbour connections 
are retained and each
site is randomly connected to another site with a probability $p$\ct{newman}. 
But there is no condition of uniqueness of random connection in NW model as in
our case.
We do not allow the possibility of two random connections emerging from 
the same site. More technically speaking, it is a regular graph
as each vertex has three incident edges. However, two of these edges are
nearest and the third is random though unique.

The tips of the fibers are now placed on 
these $N$ sites and are associated with a  
threshold strength chosen randomly from a uniform distribution. If $p$ is the
density of threshold strength of fibers, then a uniform distribution implies: 
$$p = 1~~~~~0\leq\sigma_{th}\leq1$$
$$= 0~~~~~{\rm otherwise}.$$

In the passing, let us comment that a graph as generated above resembles Bethe
lattice in the thermodynamic limit $(N \to \infty)$ \ct{dhar, bollobas, stanley}. 
The similarity arises due
to the loopless structure of both, the Bethe lattice and the above
construction as $N \to \infty$. 
Although our construction has loops, it can be shown that 
there are typically very few small loops 
as $N\to \infty$.
But there is a crucial difference between the dynamics of 
Random Field Ising Model (RFIM) studied by Dhar $et.al$ and that of RFBM. 
In the RFIM, no connection is lost in the course of dynamics whereas in RFBM,
random connections disappear as the fibers start to fail under an
external load (as explained below).

When an external force is applied, fibers with threshold smaller than the
stress generated due to the external force break. The broken fibers have to 
redistribute the stress carried by them to the nearest intact fibers.
We have implemented the following algorithm in this work:

\noindent $(i)$ The system is exposed to a total load $N\sigma$ 
such that fiber at each site $i$ 
carries a stress $\sigma(i)=\sigma$. If $\sigma_{th}(i)<\sigma(i)$ at
any site $i$, the fiber breaks and its stress is equally distributed to its 
surviving neighbours (both directly and randomly connected). 
The breaking condition is examined for each site and is 
repeated until there is no more breaking for this particular external load. 

\noindent $(ii)$ The total load is now increased from
$N\sigma$ to $N(\sigma+d\sigma)$. The new total load is then shared by 
the remaining unbroken fibers and once again the breaking condition is 
checked for each intact fiber. 

\noindent $(iii)$ It may so happen that during the failure dynamics 
a fiber encounters a situation where
all its nearest neighbours are broken. This situation can be treated in two
 ways.  One approach assumes that
the stress carried by such a fiber is lost $i.e.$, the total load is not 
conserved
\ct{kim}. On the other hand,  we study a FBM with total external
 load being strictly
 conserved.
To ensure this conservation, we propose the following: 
if a fiber encounters a situation in which one of its directly connected
neighbours (in clockwise or anticlockwise direction) is broken,
it gives its stress to the next nearest surviving neighbour in that direction.
This rule however is not applicable to the randomly connected fiber $i.e.$,
if the randomly connected neighbour of a broken fiber is already broken, 
the stress then is given only to its directly connected neighbours. 

\noindent $(iv)$ Once again,
the load is increased and the process is repeated till the complete 
breakdown of the system.
\begin{figure}
\begin{center}
\includegraphics[height=2.0in,width=2.2in]{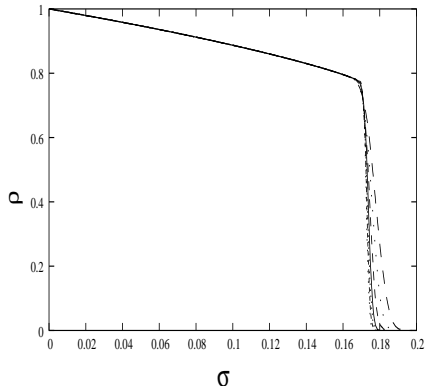}
\end{center}
\caption{The variation of $\rho$ with $\sigma$ for different values of $N$.  
Here, $N=2^{13}, 2^{14}, 2^{15}, 2^{16}, 10^{5},2^{17},
2*10^{5}, 2^{18}$. The lower $\sigma_{c}$ corresponds to a higher value of $N$.}
\end{figure}

It is clear from the above load sharing rule that even if the tips of fibers
 when placed
on a graph as mentioned before resemble a Bethe lattice, this
resemblance is valid only at time $t=0$. The broken randomly 
connected neighbour of site
$i$ will reduce the co-ordination number of site $i$ deviating the lattice from
its initial geometry. We illustrate the third point of the
algorithm with an example from Fig.~1. Let us assume that
the fiber at the site 14 is the weakest among all the fibers connected to it.
At any instant of time $t$, fiber at site
14 breaks. The additional load is shared equally among fibers at sites 11, 13
and 15. Consider a situation where at some higher load,  
fiber at site 11 also breaks; the 
load carried by 11 is equally shared only between the two 
directly connected fibers (at sites 10 and 12) if they are intact.
Otherwise, the load is given to
the next intact fiber along the direction of the broken directly 
connected fiber. Therefore effectively the long range random connection 
between 11 and 14 disappears as soon as 14 breaks.

In our numerical simulations, $d\sigma$ is taken to be 0.0001 and the average is taken over 
$10^{4}$ to $10^{5}$ configurations. The variation of
fraction of unbroken fibers $\rho$, with the applied stress is shown in Fig.~2.
Also shown is the dependence of critical stress $\sigma_c$ on N in Fig.~3.
Clearly, there exist a non-zero 
value of $\sigma_c$ for large $N$ whereas in the case of pure local
load sharing in a one-dimensional lattice, $\sigma_c$ goes to zero in the
thermodynamic limit\ct{pacheco}.
Here, $\rho$ behaves as the order parameter of the system which vanishes 
at the critical point.

\begin{figure}
\begin{center}
\includegraphics[height=2.2in,width=2.4in]{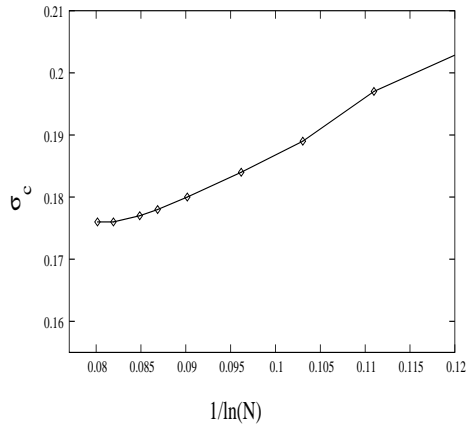}
\end{center}
\caption{The dependence of critical stress $\sigma_c$ on the system size $N$.}
\end{figure}

To determine $\sigma_{c}$ precisely, we use the standard finite size 
scaling method \ct{fisher} with the scaling assumption
\begin{equation}
\rho(\sigma,N) = N^{-a}f((\sigma-\sigma_{c})N^{\frac{1}{\nu}})
\end{equation}
where f(x) is the scaling function and the exponent $\nu$ describes the 
divergence of the correlation length $\xi$ near the critical point, $i.e.$, 
$\xi \propto |\sigma-\sigma_{c}|^{-\nu}$. 
Demanding that the order parameter $\rho$ scales as 
$\rho \propto (\sigma_{c}-\sigma)^{\beta}$, we arrive at the relation 
$a=\beta/\nu$. The plots of $\rho N^{a}$ with $\sigma$ for different $N$
crosses through a unique point $\sigma=\sigma_{c}$=0.173 with the exponent 
$a=0.5$ (See Fig.~4). 

\begin{figure}
\begin{center}
\includegraphics[height=2.2in,width=2.2in]{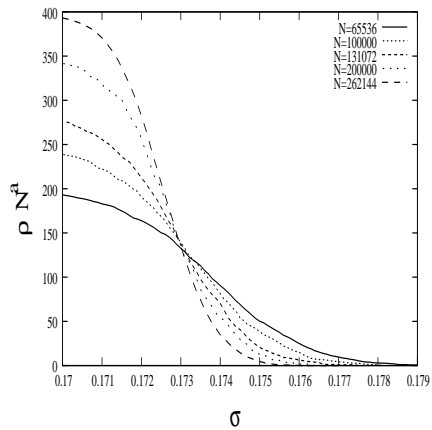}
\end{center}
\caption{Precise determination of $\sigma_{c}$ from the finite size scaling
form (1). The $\sigma_{c}$ at which
the curves intersect is 0.173. $a=0.5$.}
\end{figure}
With this value of $\sigma_{c}$ and $a$, one can again use the scaling 
relation (1) to determine the  exponent $\nu$ by making the data points to 
collapse to an almost single smooth curve near the critical point 
as displayed in Fig.~(5). The 
value of the exponent $\nu$ obtained in this manner 
is equal to unity which automatically 
gives the value of the other exponent $\beta$=0.5. 
These matches with the 
mean field exponents derived analytically\ct{dynamic}. 
It should be noted that in Fig.~(5), the  collapse of data is better for higher
system size.
The results therefore establish the existence of a sharp thermodynamic 
transition with mean field (GLS) exponents eventhough the load sharing rule
is local apart from random connections. 
\begin{figure}
\begin{center}
\includegraphics[height=2.2in,width=2.2in]{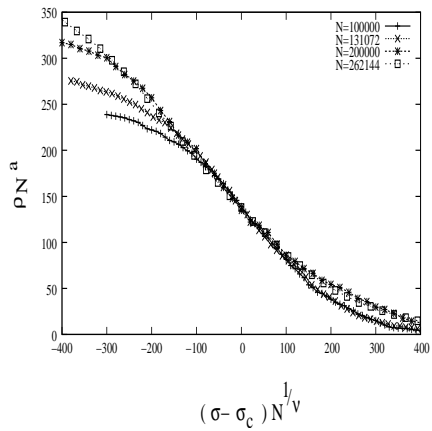}
\end{center}
\caption{Data collapse with $a$=0.5, $\nu=1$}
\end{figure}

An important quantity associated with any breakdown process  is the 
avalanche size
distribution. An avalanche is defined as the number of fibers broken between 
two successive external loadings. For RFBM with GLS, the distribution 
$D(\Delta)$ of an avalanche of size 
$\Delta$ follows a power law given as
\ct{hansen92}

\begin{equation}
D(\Delta)\propto \Delta^{-\zeta}, {\rm where~ \zeta = 5/2}.
\end{equation}

\noindent when $\Delta$ is large.

To study the avalanche size distribution in the present model, 
we have applied the weakest fiber
failure approach. In this approach, an external force is increased
quasistatically such that only the weakest surviving fiber present in the 
bundle breaks. The failure of this weakest fiber causes an avalanche
of failures of size $\Delta$. The numerical result for the distribution of 
avalanche size is shown in Fig.~6.
The exponent $\zeta$ takes the mean field value 5/2 for very large $\Delta$.
\begin{figure}
\begin{center}
\includegraphics[height=2.2in,width=2.2in]{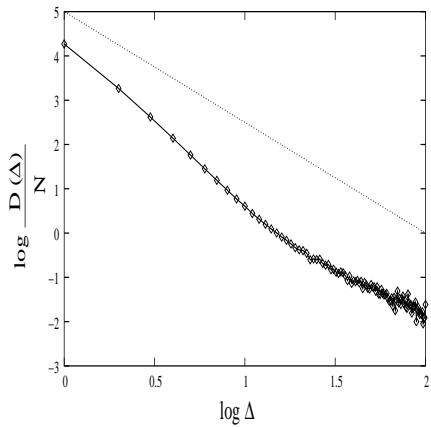}
\end{center}
\caption{Avalanche size distribution of fibers placed on a random graph.
Also drawn is a straight line with slope 5/2.
Clearly, the exponent $\xi$ takes the mean field value 5/2 for large $\Delta$.}
\end{figure}

\section{Conclusion}
We have studied a fiber bundle model where tips of fibers are placed on a 
graph with co-ordination number $3$. A local load sharing mechanism is employed for the redistribution of load of the broken fiber to the intact fibers
as described in Section II. The exact value of the critical
stress and the critical exponents $\beta$ and $\nu$ are obtained using 
finite size scaling method. These exponents correspond to the mean field 
(GLS) value as obtained in Ref.\ct{dynamic}. 
The mean field behaviour is due to 
a finite fraction of infinite range connection
which survive in the process of dynamics.

The avalanche size distribution is studied
using weakest fiber failure approach. 
We observe that the burst avalanche distribution picks up the mean field
value $\zeta=5/2$ only in the asymptotic limit ($i.e.,$ for large $\Delta$
which occurs near the critical point).
For smaller $\Delta$, the behaviour is clearly different from the
mean field limit. This suggests that the load sharing rule
affects the behaviour of $D(\Delta)$ in the small $\Delta$ region. 
The small $\Delta$-anomaly  is not seen
in  fibers placed on a complex network\ct{complex}.
 
\begin{center}
{\bf ACKNOWLEDGMENTS}
\end{center}
The authors thank B. K. Chakrabarti and S. Pradhan for useful comments.

\end{document}